\documentclass[aps,prl,twocolumn,superscriptaddress,groupedaddress]{revtex4-1} 
\usepackage{graphicx}  
\usepackage{dcolumn}   
\usepackage{bm}        
\usepackage{amssymb}   
\usepackage{amsmath}
\usepackage{transparent}
\usepackage{hyperref}

\begin{document}

\title[The Hilbert transform]{The Hilbert transform: Applications to atomic spectra}

\author{Kate A. Whittaker} 
\author{James Keaveney}
\author{Ifan G. Hughes}
\author{Charles S. Adams}
\address{Joint Quantum Centre (JQC) Durham-Newcastle, Department of Physics, Durham University,South Road, Durham, DH1 3LE, United Kingdom}

\begin{abstract}

In many areas of physics, the Kramers-Kronig (KK) relations are used to extract information about the real part of the optical response of a medium from its imaginary counterpart. In this paper we discuss an alternative but  mathematically equivalent approach based on the Hilbert transform. We apply the Hilbert transform to transmission spectra to find the group and refractive indices of a Cs vapor, and thereby demonstrate how the Hilbert transform allows indirect measurement of the refractive index, group index and group delay whilst avoiding the use of complicated experimental set ups.

\end{abstract}

\maketitle


\section{I. Introduction}

Causality is an important theme in physics and when its consequences are considered from a quantum and atomic optics perspective, relations between the real and imaginary part of the optical response can be derived. For example, the Kramers-Kronig (KK) relations link the real and imaginary parts of the optical response, which relate to the refractive index and opacity of a medium, respectively \cite{KRONIG:26,kramers1927diffusion}. The KK relations are widely used in many fields of physics and electronics \cite{Chiesa2014KKelectric}, including plasmonics \cite{Gorkunov2014KKplasmon}, and electron spectroscopy \cite{Youichi1989espec}, and find utility in applications such as light propagation, including pulse stopping \cite{Bajcsy2003pstop}, superluminal propagation \cite{Wang2000superlum,Keaveney2012SL} and quantum memories \cite{QuantumMem}.

Whereas measurements of absorption are commonplace, determination of the concomitant dispersion spectra are relatively scarce. In order to fully characterize the optical response, knowledge of both the frequency dependent refractive index $n(\omega)$ and group refractive index $n_{\mathrm{g}}(\omega)$, and the group delay \cite{Dicaire2014HTboyd} are required. This necessitates either relatively complex interferometric experiments \cite{Xiao1995,Bason2008Jamin,heterodyne_vahid} or numerics based on the KK relations involving a computationally intensive integral over all positive space of the imaginary part of the complex refractive index to determine a single frequency value of the real part \cite{Jundt2003sagnacKK}. 

Alternatively, one can exploit the fact that KK relations are a form of Hilbert transform \cite{Hall2009,Mecozzi2009}, which is a standard function incorporated into many signal processing packages, offering a significant reduction in computation time. The basic Hilbert transform can be used to quickly convert a real function to its imaginary counterpart. Although the Hilbert transform is well known in the field of communications \cite{Yang2014commHT}, signal processing \cite{Hall2009}, and has been applied to group delay measurements on fibre Bragg gratings \cite{Dicaire2014HTboyd}, it is relatively underutilized in the field of quantum and atomic optics, where measurements and predictions of the refractive index are often needed.

Here, we focus on the application of the Hilbert transform to the determination the refractive index, group index and group delay in an atomic ensemble using only transmission data. We test the validity of the method on an optical medium, atomic Cs vapor \cite{Whittaker2014}, where the real and imaginary part of the optical response are  known theoretically \cite{ElecSus}. We conclude that to the accuracy of our experimental method, the Hilbert transform can be used to reliably predicts the index and group index, and thereby provides a convenient route to obtain these quantities.
We also use the Hilbert transform to extract information about the group index, which is instrumental in the investigation of fast and slow light phenomena \cite{Boyd2009,siddonsnphotonics}.

\section{II. Theory}

\begin{figure}
\centering
\includegraphics[scale = 0.5]{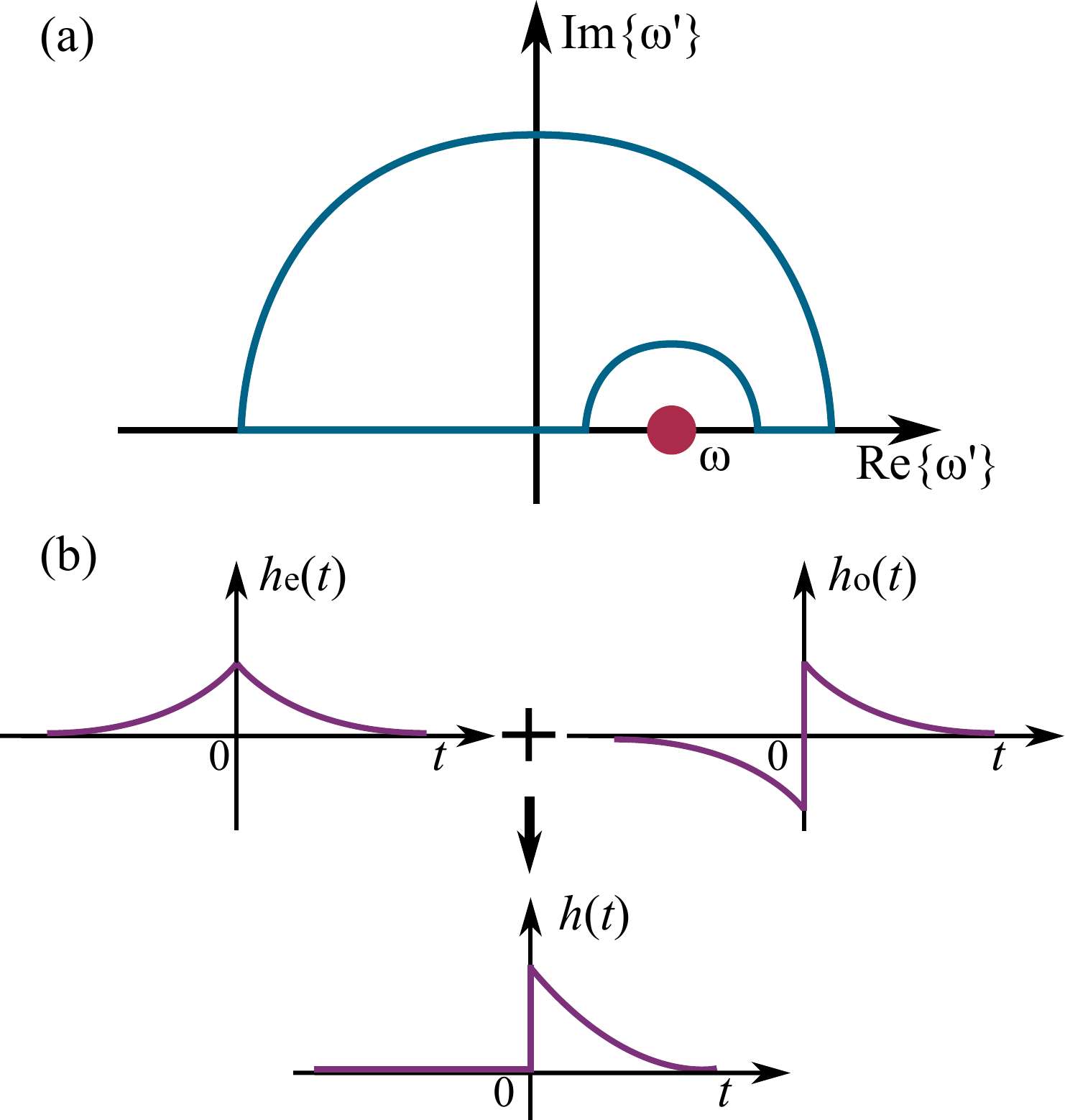}
\caption{\label{fig:contour}(a) The contour (blue line) used to derive the Kramers-Kronig relations, along with the pole at $\omega'$ = $\omega$ (red dot). To solve, the limits of the large semicircle are pushed to infinity, whilst the radius of the smaller semicircle surrounding the pole is reduced towards zero. (b) Visualization of properties of odd and even components of a causal function. The summation of odd $h_\mathrm{o}(t)$ and even $h_\mathrm{e}(t)$ components give the causal function $h(t)$ that is zero for $t < 0$}
\end{figure}

The KK relations are used to find the real part of the frequency dependent susceptibility $\chi_\mathrm{_R}(\omega)$ from the imaginary part $\chi_\mathrm{_I}(\omega')$. They can be derived by solving the following integral:
\begin{equation}\label{eqn:1KK_rel}
\chi_\mathrm{_R}(\omega) =-\frac{1}{\pi} \int_{-\infty}^{\infty}\frac{\chi_\mathrm{_I}(\omega')}{\omega'-\omega} \;\mathrm{d}\omega'.
\end{equation} 
If $\chi(\omega)$ is real and analytic in the upper half of the complex plane, a solution can easily be found by applying a well chosen contour, shown in Fig.~\ref{fig:contour}a. Since no poles lie inside the contour, Cauchy's residue theorem \cite{arfken2011mathematical} states that
\begin{equation}
\oint_{C}^{ }\frac{\omega'\chi(\omega')}{\omega'-\omega}\;\mathrm{d}\omega' = 0.
\end{equation}
This eventually results in the well known KK relation (for a more in depth derivation see e.g. \cite{boyd2003nonlinear}):
\begin{equation}
\chi_\mathrm{_R}(\omega) = 1+ \frac{2}{\pi}\int_{0}^{\infty}\frac{\omega'\chi_\mathrm{_I}(\omega')}{\omega'^2-\omega^2}\;\mathrm{d}\omega'.
\end{equation}
The KK relations can also be derived using analysis in the time domain, avoiding contour integration \cite{KK_tdomderiv,Hall2009}. For this we need to utilize 3 main properties of causal functions:
\begin{enumerate}
\item Any causal function $h(t)$ can be expressed as the sum of even ($h_{\mathrm{e}}(t)$) and odd ($h_{\mathrm{o}}(t)$) functions, the sum of which is 0 when $t < 0$.
\item The Fourier transform of $h_{\mathrm{e}}(t)$ is purely real.
\item The Fourier transform of $h_{\mathrm{o}}(t)$ is purely imaginary.
\end{enumerate}
For clarity, property (1) is illustrated in Fig.~\ref{fig:contour}b. In addition to this, we will also need to use the signum function $\mathrm{sgn}(t)$. It converts an even function to an odd function and vice versa, and is defined as:
\begin{equation}
\mathrm{sgn}(t) = \bigg\{  \begin{array}{r@{\quad}cr} 
1 & \mathrm{if} & t > 0, \\  
-1 &  \mathrm{if} & t < 0.  
\end{array}
\end{equation}
If we consider a real and causal function $h(t)$, composed of a sum of even and odd functions $h_{\mathrm{e}}(t)$ and $h_{\mathrm{o}}(t)$, then following principles (2) and (3), the Fourier transform of $h(t)$, denoted as $H(\omega)$, is composed of a purely real component $\mathcal{F}\left [ h_{\mathrm{e}}(t) \right]$ and a purely imaginary component $\mathcal{F}\left [ h_{\mathrm{o}}(t) \right]$. Using the signum function and using property (1), $h_{\mathrm{e}}(t)$ can be expressed in terms of $h_{\mathrm{o}}(t)$:
\begin{equation}
h(t) = h_{\mathrm{o}}(t) + h_{\mathrm{e}}(t) = h_{\mathrm{o}}(t) + \mathrm{sgn}(t)h_{\mathrm{o}}(t).
\end{equation}
The Fourier transform of $h(t)$, $H(\omega)$, gives us a link between the real and imaginary parts of $H(\omega)$. To understand this further, we note that multiplication in the time domain is equivalent to a convolution in the frequency domain. Hence, the Fourier transform of $h(t)$ becomes:
\begin{equation}\label{eqn:FTHT}
\mathcal{F}\left[h(t)\right] = \mathcal{F}\left[h_{\mathrm{o}}(t)\right] + \mathcal{F}\left[\mathrm{sgn}(t)\right]\ast \mathcal{F}\left[h_{\mathrm{o}}(t)\right],
\end{equation}
where $\ast$  denotes convolution. The Fourier transform of a signum function is $-i/\pi\omega$ \cite{bracewell2000fourier}, and so the second part of equation \ref{eqn:FTHT} becomes a convolution between $H_{\mathrm{o}}(\omega)$ and the kernel $1/\pi\omega$, better known as the Hilbert transform \cite{arfken2011mathematical}. The Hilbert transform is well known and is expressed as
\begin{equation}
\hat{H}(\omega) = \frac{1}{\pi}\int_{-\infty}^{\infty}\frac{H(\omega')}{\omega-\omega'} \; \mathrm{d}\omega',
\end{equation}
where the hat in $\hat{H}(\omega)$ denotes that a Hilbert transform has been performed, a notation we will use throughout the rest of this paper. We can now express $H(\omega)$ in simpler terms:
\begin{equation}\label{H_equiv}
H(\omega) = H_{\mathrm{o}}(\omega) -\mathrm{i}\hat{H_{\mathrm{o}}}(\omega).
\end{equation}
Recalling properties (2) and (3) of causal functions defined earlier, it is clear that equation \ref{H_equiv} shows us that the imaginary part of $H(\omega)$ can be found by taking the real part of $H(\omega)$. Expressing this relation in terms of the susceptibility yields:
\begin{equation}\label{chi_H}
\chi_\mathrm{_R}(\omega) = \hat{\chi}_\mathrm{_I}(\omega).
\end{equation}
The result is useful for atom opticians and anyone who needs to convert an imaginary signal to its real counterpart because the Hilbert transform is a standard function in many signal processing toolboxes e.g. in Python, MATLAB and C. Algorithms are available to approximate the solution, taking only a fraction of a ms to calculate as compared to tens of seconds for an equivalent Kramers-Kronig computation.

\section{III. Experimental Methods}

\begin{figure}[t]
\centering
\includegraphics[scale = 0.37]{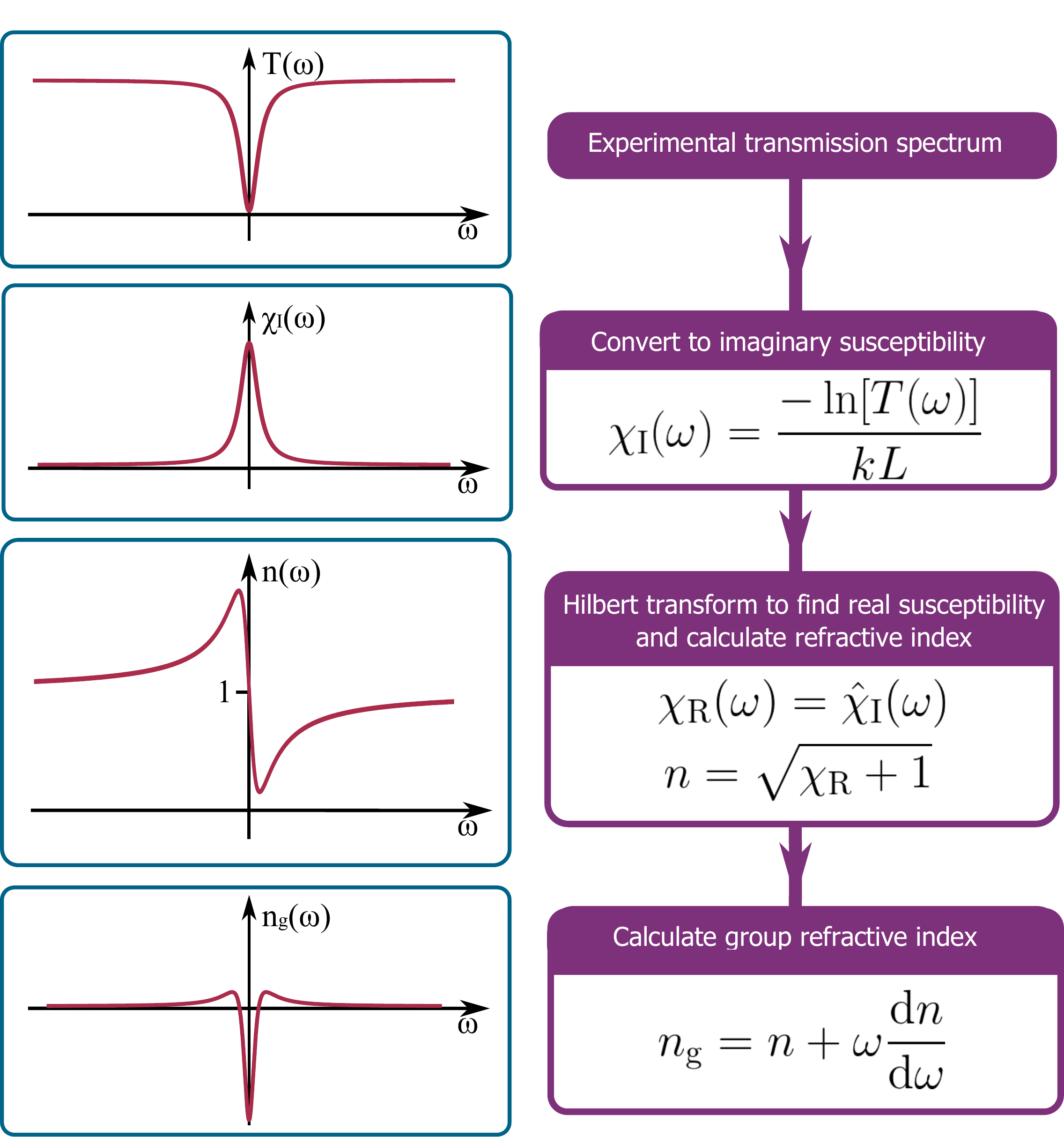}
\caption{\label{fig:process}Process for calculating Hilbert transform from transmission spectra. A transmission spectrum is taken then converted to the imaginary part of the susceptibility. The imaginary susceptibility is then Hilbert transformed and converted to the refractive index. The group index is then calculated from $n(\omega)$.}
\end{figure}

In order to verify that the Hilbert transform generates accurate line shapes for $\chi_\mathrm{_R}(\omega)$, we compare Hilbert transformed absorption spectra to the expected refractive index from an analytical model we have developed, based on the model described in \cite{ElecSus}. The analytical model calculates the full susceptibility, then takes the imaginary part to fit absorption spectra. To test the Hilbert transform, we take the real part of the susceptibility and compare it to the Hilbert transform of the absorption spectrum. We then compare it to an equivalent KK calculation, and compare computation times.

The transmission spectra used in this paper are from a nm scale vapor cell, an ultra thin alkali vapor cell with a thickness ranging from 2 $\mu$m to 30 nm \cite{Sarkisyan2001ETC}. These spectra feature peaks that have been narrowed by motional effects inside the cell (Dicke narrowing), discussed in our earlier papers \cite{Keaveney2012CLS,Keaveney2012SL}. We perform single beam transmission spectroscopy on the Cs D1 line at a cell thickness of 670 $\pm$ 10 nm at a temperature of 220 $\pm$ 5 $^\circ$C. The experimental set up is again detailed in our previous publications. The process used to extract the refractive index using the Hilbert transform is illustrated in Fig.~\ref{fig:process}. The transmission spectra $T(\omega)$ are converted to $\chi_\mathrm{_I}(\omega)$ using $\chi_\mathrm{_I}(\omega) = -\mathrm{ln}(T(\omega)/kL)$ where $k = 2\pi/\lambda$ is the wave vector, and $L$ the cell thickness. A Hilbert transform of $\chi_\mathrm{_I}(\omega)$ is taken, and the refractive index is found by taking $n(\omega) = \sqrt{1+\chi(\omega)}$. We then extract the group index $n_{\mathrm{g}}(\omega)$ from the Hilbert transformed spectra using 
\begin{equation}
n_\mathrm{g}(\omega) = n + \omega\frac{\mathrm{d}n}{\mathrm{d}\omega}.
\end{equation}
Theoretical spectra of $n(\omega)$ are generated using a model of the complex susceptibility which takes the Doppler broadened transitions and applies self-broadening \cite{LeeSelfB}, Dicke narrowing \cite{Sarkisyan2004dicke,dutier2003dickeN}, atom-surface interactions \cite{Whittaker2014} and reflectivity effects \cite{Dutier2003reflect}. The fit compares the transmission line shape to the theoretical line shape from $\chi_\mathrm{_I}(\omega)$. Since the full susceptibility is calculated, we can therefore extract the corresponding $\chi_\mathrm{_R}(\omega)$ from the fitting function.

\section{IV. Results and Discussion}

\begin{figure*}[t]
\centering
\includegraphics[scale = 0.39]{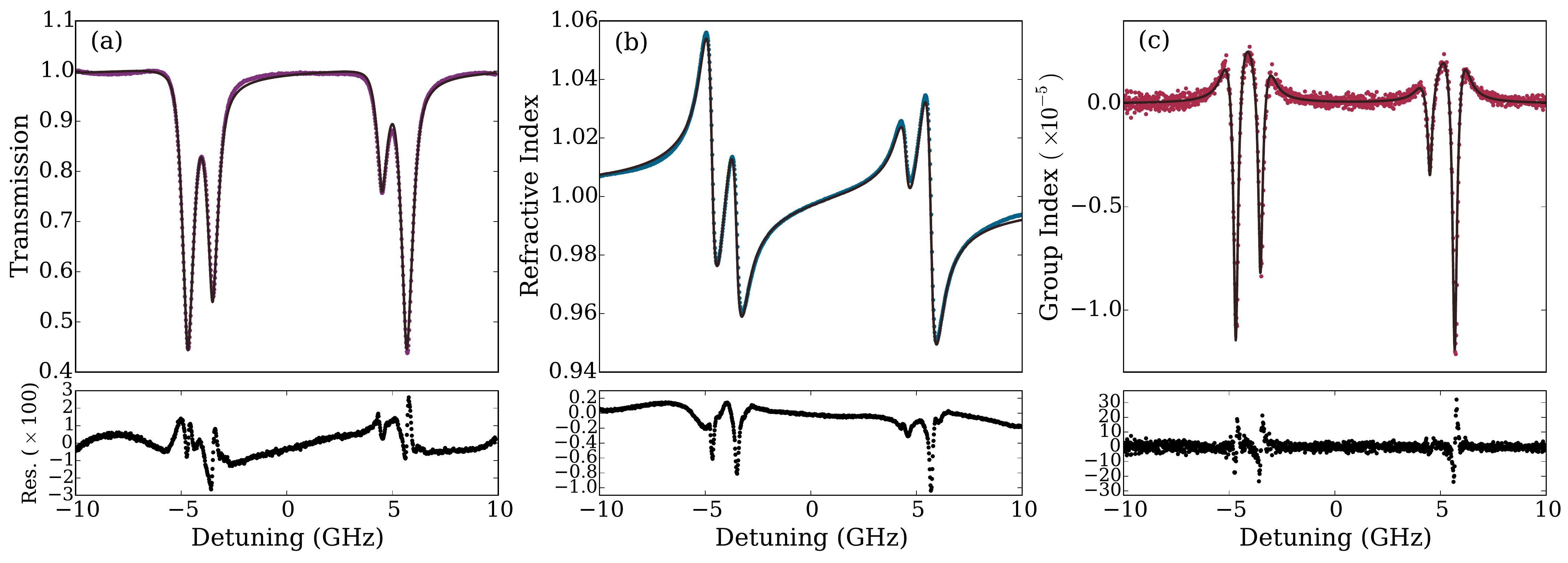}
\caption{\label{fig:trans_comp}(a): Transmission data (purple points) and theoretical fit (black line) for a Cs vapor with thickness 670 $\pm$ 10~nm and temperature 220 $\pm$ 5~$^\circ$C in the vicinity of the D1 line. (b): Refractive index inferred from the measured transmission data using the Hilbert transform (blue points) compared to theoretical expectations (black line) from our susceptibility model. (c): Group refractive index determined from the Hilbert transformed spectra (red points) and theoretical group index calculated from modified form of  \cite{ElecSus}. Below each panel are the residuals between theory and measurements using the Hilbert transform method. The residual are within the uncertainties related to laser frequency and power fluctuations.}
\end{figure*}

Fig.~\ref{fig:trans_comp} shows a comparison between theory and experiment to test the performance of the Hilbert transform. Experimental data (purple points in Fig.~\ref{fig:trans_comp}a) were taken and fitted to the model (black line); the fit has a normalised root mean squared error (NRMSE) \cite{hughes2010measurements} of 1.2\%. The data were then subjected to the treatment outlined in Fig.~\ref{fig:process}. We can see from Fig.~\ref{fig:trans_comp}b that there is excellent agreement between the fitted $n(\omega)$ (black line) and the $n(\omega)$ found from a Hilbert transform of the experimental absorption signal (blue points), with a NRMSE of 0.7\%. It should be noted that the absorption spectrum needs to be padded smoothly down to zero at either end for two reasons. Firstly, this ensures that $\chi_\mathrm{_I}(\omega)$ reaches zero at the edges of the spectra. Secondly, padding the array to a length that is a power of two vastly decreases the time required for computations, as computational Hilbert transformations use Fourier transforms. This results in calculation times of less than 0.5 ms, as calculated on a single core on a computer with an Intel Core i3 3.3 GHz processor.

For comparison, we created a simple code to calculate the refractive index using the Kramers-Kronig transform. The code is simplistic, using the a trapezium rule integration routine to calculate the integral in equation \ref{eqn:1KK_rel}. It transforms a Lorentzian mimicking a typical absorption profile to a dispersive profile, over the same number of points as the Hilbert transform shown in Fig.~\ref{fig:trans_comp}b. The code takes 11 s to calculate a refractive index profile, using the same computer that was used to time the Hilbert transform. This makes the KK calculation~$10^4$ times slower than the Hilbert transform and renders it inappropriate for possible in situ monitoring of the refractive index.

Fig.~\ref{fig:trans_comp}c shows the group index calculations performed, where the $n_{\mathrm{g}}(\omega)$ calculated from the theoretical $n(\omega)$ is compared to that calculated from the Hilbert transformed spectrum. The results are in excellent agreement with theoretical calculations. The agreement between experiment and theory is excellent, with a NRMSE of 2.6\%. This demonstrates that very reliable spectra for the group index can be extracted using the Hilbert transform. Additionally, the largest group index measured in Fig.~\ref{fig:trans_comp} is $-(1.21 \pm 0.03)\times\ 10^5$, a negative group index even larger than the previous record set in \cite{opacity_sat}.

\begin{figure}
\includegraphics[scale = 0.65]{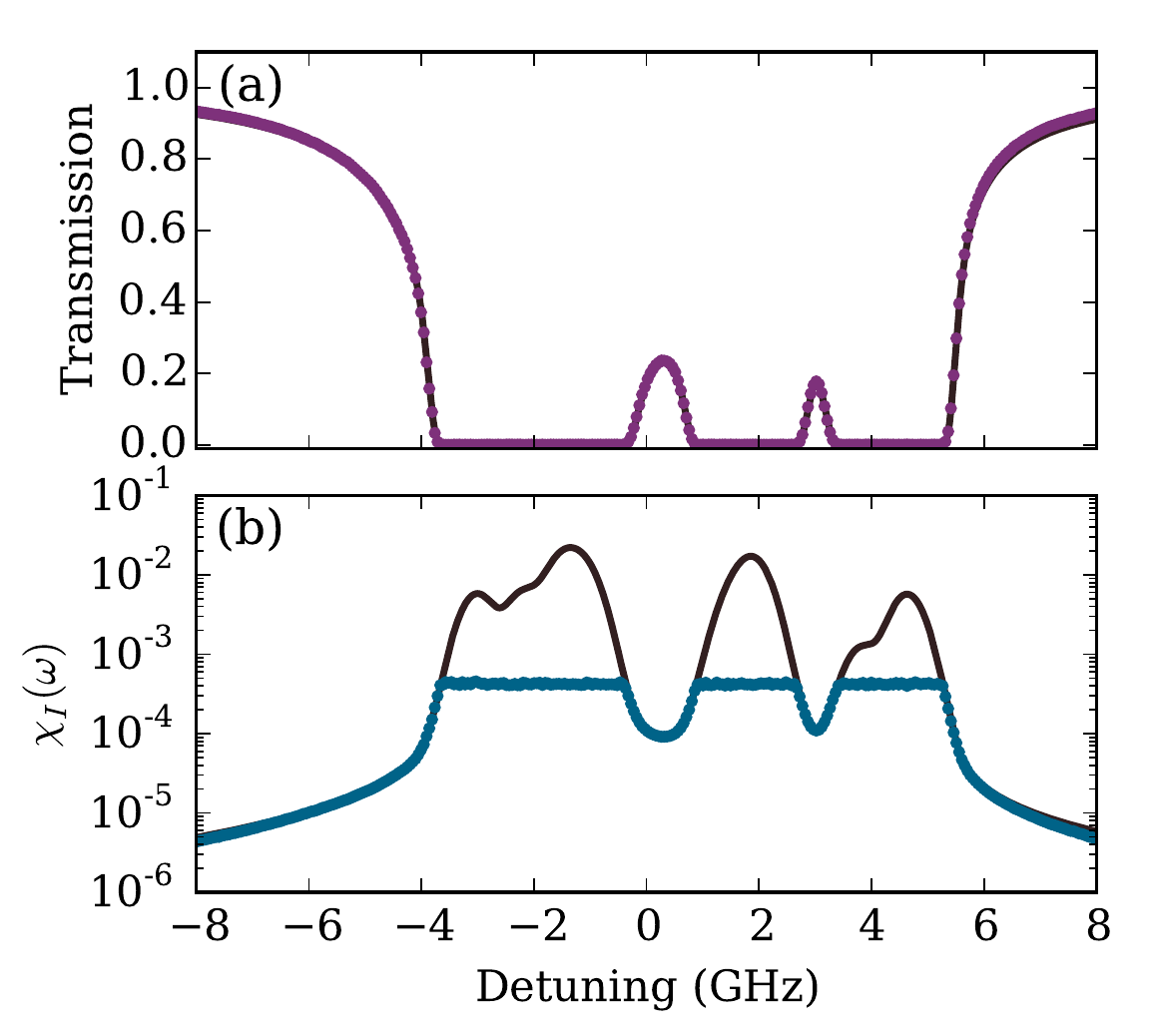}
\caption{\label{fig:opt_sat} A comparison of experimental measurements and theoretical calculations of transmission line shapes for the Rb D1 line, panel (a), and the imaginary susceptibility $\chi_\mathrm{_I}(\omega)$, panel (b), for an optically thick vapor in a 2 mm cell at T = 179$^\circ$C. Panel (a) shows the experimental (purple points) and theoretical transmission which appear to have good agreement and have a maximum expected theoretical optical density of 345. Panel (b) shows that the full $\chi_\mathrm{_I}(\omega)$ (modeled using \cite{ElecSus}, black line) cannot be determined using current detection methods (blue points).
}
\end{figure}

A limitation of this technique and KK calculations is that they cannot be applied to vapors that are optically thick. In order to extract $\chi_\mathrm{_I}(\omega)$ in optically thick vapors an excellent sensitivity and signal to noise ratio is needed - detection methods need to be sensitive to immeasurably small changes (according to values calculated using the model in \cite{ElecSus}), currently beyond what is experimentally possible.

As an example of this limitation, Fig.~\ref{fig:opt_sat} shows a comparison of the theoretical (black lines) and experimental \cite{footnote} (colored points) transmission line shapes (a) and the corresponding imaginary susceptibilities (b) for an optically thick vapor. Theoretical line shapes were generated using the model in \cite{ElecSus}. The transmission line shape measured (purple points) in panel (a) appears to match well with theoretical expectations. However, when the calculated susceptibility is plotted in a logarithmic scale in panel (b) it is clear that the experimental values (blue points) do not have enough signal resolution to capture all features of $\chi_\mathrm{_I}$ that are apparent in the theoretical susceptibility due to the noise limit of the detector. Without the full susceptibility, the Hilbert transformed $n$ cannot be calculated accurately. Hence, neither the Hilbert transform nor KK relations can be applied to infer the refractive index of an optically thick sample. However, by making the sample thinner one can move into a regime where the vapor will never become optically thick, as demonstrated in \cite{opacity_sat}.

\section{V. Conclusion}

We have demonstrated that the Hilbert transform is much faster than direct computation of the KK relations, whilst still producing results that match up well with theoretical predictions. It can be applied to many situations, providing the vapor is not optically thick. Knowledge of the refractive index can be used to calculate the group index, which find use in pulse propagation experiments and in calculating the group delay \cite{Dicaire2014HTboyd}. One could also expect that when correctly paired with appropriate data acquisition hardware, the transform should be fast enough to allow in situ monitoring of the refractive index. 

Applications could expand past the scope of pulse propagation and slow light applications, where refractive index measurements may not be necessary enough to build a complex experiment but would still be beneficial to give a deeper insight into the relevant physical processes. This could include cases when complex interactions that involve the refractive index occur, for example when investigating the alteration of the atomic line shape by etalon effects inside a vapor cell \cite{Dutier2003reflect} or to gain a fuller picture of parity non-conserving effects in transition metal vapors \cite{Cronin_thall}. Another potential use is as a diagnostic tool for testing fitting routines; the difference in dispersive spectra is clearer on visual inspection than the differences in a Voight line shape. A final possibility is to use the Hilbert transform to test existing methods of measuring the refractive index, for example the dispersion spectra in \cite{LeeAbsDisp}. We hope that this will prove to be a useful tool in the field of quantum and atomic optics, transforming measurement of the refractive index from a laborious complicated task into one that is simple and fast.

\section{Acknowledgments}

The authors would like to thank D. Sarkisyan and A. Sargsyan at the National Academy of Science, Armenia for production of the nanocells used in this paper. We also thank M. A. Zentile for providing the data for an optically thick cell. We acknowledge financial support from Durham University and the ESPRC (Grant EP/L023024/1). The data presented in this paper are available on request.

\bibliography{hilbert_bib}
\bibliographystyle{apsrev}

\end{document}